\documentstyle[preprint,aps,]{revtex}

\begin{document}
\newcommand{\bea}{\begin{eqnarray}}
\newcommand{\eea}{\end{eqnarray}}
\newcommand{\beq}{\begin{equation}}
\newcommand{\eeq}{\end{equation}}

\preprint{\begin{tabular}{r} KAIST-TH 99/05 \\ hep-ph/9907244 \end{tabular}}


\title{String theoretic axion coupling and the evolution of cosmic structures}
\author{Kiwoon Choi${}^{(a)}$,
        Jai-chan Hwang${}^{(b)}$
        and Kyu Wan Hwang${}^{(a)}$}
\address{${}^{(a)}$
         Department of Physics, Korea Advanced Institute of Science
	 and Technology, Taejon 305-701, Korea\\
         ${}^{(b)}$
         Department of Astronomy and Atmospheric Sciences,
         Kyungpook National University, Taegu, Korea 
	 }
\maketitle

\begin{abstract}

We examine the effects of the axion coupling
to $R\tilde{R}$ on the evolution of cosmic structures.
It is shown that the evolutions of
the scalar- and vector-type perturbations  are not affected by this
axion coupling. However the axion coupling causes an asymmetric evolution
of  the two polarization states of 
the tensor-type perturbation, which may lead to  a sizable
polarization asymmetry in the cosmological gravitational wave 
if inflation involves a period in which the axion coupling is important.
The  polarization asymmetry produced during inflation are conserved
over the subsequent evolution as long as the scales remain
in the large-scale limit, and thus this may lead to an observable
trace in the cosmic microwave background radiation.

\end{abstract}

\noindent

String or $M$-theory has been received much attention as the best candidate 
for the unified theory of all fundamental forces \cite{string}.
In view of that its typical energy scale 
is too high to be probed by laboratory experiments\footnote{
Recently, it has been noted that some string theories
allow the Kaluza-Klein scale or 
the string scale to be far below the Planck scale,
even as low as TeV \cite{seehowever}.},
cosmology can be one of the best testing grounds for
string or $M$-theory.
In this regard, it is important to search for a realistic
inflation model within the framework of string or $M$-theory.
The next step  would be to find out whether this inflation model
allows a successful structure formation. As is well known, due to inflation
microscopic quantum fluctuations can be magnified to macroscopic 
classical structures
which can evolve into the large-scale structures we observe today
such as the galaxy distribution and the temperature fluctuations in the cosmic
microwave background radiation (CMBR).
According to this paradigm,
the amplitude and spectrum of the large angular scale fluctuations 
in the CMBR can work as a window 
to the early universe, and particularly to  the inflation era.
Such  studies  in the context of low-energy effective action of string
theories  have  been done before \cite{pre-big-bang},
some of which include the effects of higher order 
corrections in the expansion in either   the inverse string tension
or the string coupling
\cite{string-action,Antoniadis}. 

One of few model-independent predictions of string or $M$-theory
is the existence of axions which couple to 
$F\tilde{F}=\eta^{abcd}F_{ab} F_{cd}$ and/or to 
$R \tilde R \equiv \eta^{abcd} R_{ab}^{\;\;\;\;ef} R_{cdef}$
\cite{stringaxion,Maxion} where $\eta^{abcd}$ is a (totally antisymmetric)
Levi-Civita tensor density.
It is rather easy to see that for spatially homogeneous and isotropic background
the axion coupling to $F\tilde{F}$ affects {\it neither} the evolution
of background {\it nor} the evolution of perturbations in linear approximation
\cite{ram}.
The axion coupling to $R\tilde{R}$ 
does not affect the evolution of  background also,
however it can affect the evolution of perturbations in the early 
universe \cite{Lue}.

Compared to the Ricci scalar curvature term,
$R\tilde{R}$ is higher order
in the dimensionful gravitational coupling $\kappa^2=8\pi G_N$.
Although  suppressed by $\kappa^2$,
it may lead to  sizable effects in some string inflation models
(pre-big-bang)
which encounter a curvature singularity 
when the cosmological evolution is 
governed by the lowest order effective action
\cite{pre-big-bang}.
In such inflation models, it is expected \cite{Antoniadis}
that higher dimensional terms of curvature tensor
regulate the curvature singularity to the period of large but finite
curvature $\sim 1/\kappa^2$ which would smoothly
evolve into the standard radiation dominated flat universe.
Clearly in the high curvature period,
higher dimensional terms of curvature tensor
such as the axion coupling to $R\tilde{R}$ and/or
the moduli coupling to
the Gauss-Bonnet combination \cite{string-action}
can be important as much as the Ricci scalar term
\cite{cosmo-appl,Gasperini,Kawai,GB}.

Recently, authors of \cite{Lue} considered the effect of axion coupled 
$R \tilde R$ term on gravitational waves.
They derived the gravitational wave equation in Minkowsky background
and showed that $R \tilde R$ coupling term leads to asymmetric equations
for the left and right handed polarization states of graviational waves.
In this paper, we wish to generalize the study of \cite{Lue}
by discussing  the effects of the axion coupling
on the evolution of cosmological 
structures of all types in the context of evolving cosmological background. 
It turns out that the evolutions of
scalar- and vector-type structures are not affected by the axion coupling.
However, as noted in \cite{Lue} it causes an asymmetric evolution of 
the two polarization states of tensor-type perturbation.
This  may lead to a sizable polarization asymmetry in tensor-type
perturbation if inflation involves a period
during which the axion coupling is important, such as the
high curvature period in pre-big-bang model.
Moreover, in the large-scale limit, the amplitudes of both 
polarization states are  seperately conserved as can be seen in 
Eq. (\ref{GW-LS-sol}).
Thus once the  polarization  asymmetry in tensor-type perturbation was 
generated during inflation, it will be preserved over the subsequent 
evolution stage as long as the scale remains in the large-scale limit.
This then  may lead to an observable trace in the large angular scale 
polarization asymmetry in the CMBR \cite{Lue}.

Our starting point is an effective action including the axion coupling
to $R\tilde{R}$, which may correspond to the  low energy effective
action of string or $M$-theory \cite{string}:
\bea
   S = \int d^4 x \sqrt{-g} \Big[ {1 \over 2 \kappa^2} R
       - {1 \over 2} \omega (\phi) \phi^{,c} \phi_{,c} - V (\phi)
       + {1 \over 8} \nu (\phi) R \tilde R + L_{m} \Big],
   \label{action}
\eea
where $\kappa^2=8\pi G_N$ is the reduced Newton constant,
$R$ is the scalar curvature, $V(\phi)$, $\omega(\phi)$ 
and $\nu(\phi)$ are some generic functions\footnote{
Here we use the effective action written in the Einstein frame.
The axion-dependence of the  K\"ahler metric $\omega$ and 
the potential $V$ can arise from nonperturbative effects in string theory.}
of the axion field $\phi$,
$R \tilde R \equiv \eta^{abcd} R_{ab}^{\;\;\;\;ef} R_{cdef}$, and
$L_m$ is the Lagrangian including the fields other than
the axion field $\phi$.
Here we are interested in how the evolution
of axion background affects the evolution of cosmic structures
through the coupling to $R\tilde{R}$, and thus
other  axion couplings, e.g. the axion coupling to
moduli, are ignored for the sake of simplicity.
Generically string or $M$-theory  predicts 
many axions \cite{stringaxion,Maxion}, and then
$\phi$ may be identified as the axion combination which couples
to $R\tilde{R}$.
The gravitational field equation is given by
\bea
   & & R_{ab} - {1 \over 2} R g_{ab} 
       = \kappa^2 \Big[ \omega (\phi) ( \phi_{,a} \phi_{,b} 
       - {1 \over 2} \phi^{,c} \phi_{,c} g_{ab} )
       - V(\phi) g_{ab} + \tilde T_{ab} + T^{(m)}_{ab} \Big],
   \nonumber \\      
   & & \tilde T_{ab} \equiv \eta_{(a}^{\;\;\;\; cde} 
       \big( R^f_{\;\;\; b) cd} \nu_{,e;f} - 2 R_{b)c;d} \nu_{,e} \big),
   \label{GFE}
\eea
where $X_{,a}$ and $X_{;a}$ denote the normal and covariant derivatives 
of $X$ respectively, $X_{(ab)} \equiv {1 \over 2} (X_{ab} + X_{ba})$, and
$T^{(m)}_{ab}$ is the additional energy-momentum tensor arising from $L_m$
which would include the contribution from cosmic fluid.

As the spacetime metric, we consider a spatially flat, homogeneous, 
and isotropic background including the most general form of
space-time dependent perturbations:
\bea
   d s^2 
   &=& - a^2 \left( 1 + 2 \alpha \right) d \eta^2
       - a^2 \left( \beta_{,\alpha} + B_\alpha \right) d \eta d x^\alpha
   \nonumber \\
   & & + a^2 \Big[ \delta_{\alpha\beta} \left( 1 + 2 \varphi \right)
       + 2 \gamma_{,\alpha\beta} + 2 C_{\alpha,\beta}
       + 2 C_{\alpha\beta} \Big] d x^\alpha d x^\beta,
   \label{metric}
\eea
where $a(t)$ is the cosmic scale factor with $dt \equiv a d \eta$.
Here $\alpha ({\bf x}, t)$, $\beta ({\bf x}, t)$, $\varphi ({\bf x}, t)$
and $\gamma ({\bf x}, t)$ characterize the scalar-type perturbations,
$B_\alpha ({\bf x}, t)$ and $C_\alpha ({\bf x}, t)$ are the transverse
vector-type perturbation, and finally
$C_{\alpha\beta} ({\bf x}, t)$ stands for  the tracefree,
transverse tensor-type perturbation.
The spatial indices   are based on the metric $\delta_{\alpha\beta}$.
We also decompose the energy-momentum tensor and the axion field as
\bea
T^a_b ({\bf x}, t) = \bar T^a_b (t) + \delta T^a_b ({\bf x}, t),
\quad
\phi ({\bf x}, t) = \bar \phi (t) + \delta \phi ({\bf x}, t).
\eea
In the following, the overbar
in background configuration
will be omitted unless necessary.
Since the three types of perturbations decouple from each other due to 
the symmetry of the background and also the assumed  linearity of the 
structures, we can handle them individually.

It is well known that $\nu(\phi)R \tilde R$
does not affect the equation for  background \cite{Antoniadis}.
After some calculation we can show that for the scalar-type perturbation 
we have $\tilde T_{ab} = 0$, and as a result
the axion coupling does not affect the evolution of the scalar-type
perturbation as well.
This can be explained by that we cannot form a scalar
($\tilde T_{00}$) or a vector ($\tilde T_{0\alpha}$)
or a symmetric tensor ($\tilde T_{\alpha\beta}$)
which contains $\eta^{abcd}$ 
with the derivatives of scalar and $\delta_{\alpha\beta}$ only. 
Thus the evolution of scalar-type perturbation
is the same as that in the case without the axion coupling
which has been studied in \cite{GB}.

We have nontrivial contributions from the axion coupling
$\nu(\phi) R \tilde R$ only in the equations for vector-
and tensor-type perturbations.  
({}For explicit calculations, Appendices
of \cite{Rab-GW,Rab-rot} will be useful.)
{}For the vector-type perturbation, Eq. (\ref{GFE}) leads to 
\bea
   & & {k^2 \over 2 a^3} ( a \Psi_\alpha - \kappa^2 \dot \nu
       \epsilon_\alpha^{\;\;\;\gamma\delta} \Psi_{\gamma,\delta}) 
       = \kappa^2 \delta T^{(m)0}_{\;\;\;\;\;\;\alpha},
   \label{rot-E1} \\
   & & {1 \over a} 
       {\partial \over \partial t} \Big[ a ( a \Psi_{(\alpha} 
       - \kappa^2 \dot \nu \Psi_{\gamma,\delta} 
       \epsilon_{(\alpha}^{\;\;\;\;\gamma\delta})_{,\beta)} \Big]
       = \kappa^2 \delta T^{(m)}_{\alpha\beta}.
   \label{rot-E2}
\eea
where we have 
introduced
\bea
   \epsilon^{\alpha\beta\gamma}
   \equiv a^4 \bar \eta^{0\alpha\beta\gamma},
        \quad
       \Psi_\alpha \equiv B_\alpha + a \dot C_\alpha
\eea
with the overdot denoting the derivative w.r.t. $t$.
When expressed in terms of the 
following notation based on the vector-type harmonic function
$Y^{(v)}_\alpha$ 
with $Y^{(v)}_{\alpha\beta} \equiv - k^{-1} Y^{(v)}_{(\alpha,\beta)}$
\cite{Rab-rot},
\bea
   \delta T^{(m)0}_{\;\;\;\;\;\;\alpha} 
   \equiv ( \mu + p) v_\omega Y^{(v)}_\alpha, 
       \quad
       \delta T^{(m)}_{\alpha\beta} 
       \equiv \pi^{(v)} Y^{(v)}_{\alpha\beta},
\eea
Eqs. (\ref{rot-E1}) and (\ref{rot-E2}) lead to
\bea
   {1 \over a^4} {\partial \over \partial t} 
       \Big[ a^4 \left( \mu + p \right) \times v_\omega \Big] 
       = - {k \over 2a} \pi^{(v)},
   \label{rot-eq}
\eea
where 
$v_\omega$ is related to the rotational velocity of  vector-type
fluid perturbations and $\pi^{(v)}$ is the anisotropic stress 
of the fluid which can 
work as the sink or source of the rotation \cite{Rab-rot}. 
If we ignore the anisotropic stress, the angular-momentum of
the {\it fluid} whose energy momentum tensor  is given
by $T^{(m)}_{ab}$
is conserved as
\bea
   a^4 \left( \mu + p \right) \times v_\omega ({\bf x}, t) \sim L({\bf x}).
   \label{rot-sol}
\eea
Therefore, the presence of the axion coupling does {\it not}
affect the rotational
type perturbation of the fluid component
which is described again by the angular 
momentum conservation, although the evolution of the associated metric
components $\Psi_{\alpha}$ is affected by the axion coupling
as is apparent in Eq. (\ref{rot-E1}).

In fact, it is the tensor-type perturbation that may receive
an observable impact from the axion coupling.
Eq. (\ref{GFE}) gives the following 
equation for the evolution of tensor-type perturbation:
\bea
   D_{\alpha\beta} - {2 \kappa^2 \over a}  
       \epsilon_{(\alpha}^{\;\;\;\;\gamma\delta}
       \Big[ ( \ddot \nu - H \dot \nu ) \dot C_{\beta)\gamma}
       + \dot \nu D_{\beta)\gamma} \Big]_{,\delta}
       = \kappa^2 \delta T^{(m)}_{\alpha\beta},
\label{GW-eq1}
\eea
where
\bea
   D_{\alpha\beta} \equiv \ddot C_{\alpha\beta}
       + 3 H \dot C_{\alpha\beta} - {1\over a^2} \Delta C_{\alpha\beta}.
\eea
(Here $\Delta$ denotes the spatial Laplacian.)
Similar  equation of gravitational wave in the presence of axion coupling 
was derived in \cite{Lue}, but only for the Minkowski background.
Let us expand the tensor perturbation \cite{GGT-GW} as
\bea
   C_{\alpha\beta} ({\bf x}, t)
       \equiv \sqrt{\rm Vol} \int {d^3 k \over ( 2 \pi)^3}
       \sum_\ell e^{(\ell)}_{\alpha\beta} ({\bf k})
       h_{\ell {\bf k}} (t) e^{i {\bf k} \cdot {\bf x}},
\eea
where $e^{(\ell)}_{\alpha\beta}$ is the circular polarization tensor 
($\ell = L, R$) with the property 
$i k_{\gamma}\epsilon_\alpha^{\;\;\; \gamma\delta} e^{(\ell)}_{\beta\delta} 
= k \lambda_\ell e^{(\ell)}_{\alpha\beta}$ ($\lambda_L = - 1$
and $\lambda_R = +1$) where $k \equiv |{\bf k}|$.
Then ignoring the anisotropic stress of the additional fluid,
Eq. (\ref{GW-eq1}) becomes 
\bea
   {1\over a^3 ( 1 + \lambda_\ell \kappa^2 \dot \nu k/a ) }
       {\partial \over \partial t} 
       \Big[ a^3 ( 1 + \lambda_\ell \kappa^2 \dot \nu k/a ) 
       \dot h_{\ell {\bf k}} \Big]
       + {k^2 \over a^2} h_{\ell {\bf k}} = 0.
   \label{GW-eq}
\eea
If we set $\psi_{\ell {\bf k}} \equiv z_\ell h_{\ell {\bf k}}$
with $z_\ell \equiv a \sqrt{ 1 + \lambda_\ell \kappa^2 \dot \nu k/a }$,
we can easily show 
\bea
   \psi_{\ell {\bf k}}^{\prime\prime} + \left( k^2 
       - {z_\ell^{\prime\prime} / z_\ell} \right) \psi_{\ell {\bf k}} = 0,
   \label{GW-eq2}
\eea
where the prime denotes the derivative w.r.t. the conformal time $\eta$.  

In many prototype  inflation models,
$z_\ell^{\prime\prime}/z_\ell = n_\ell/\eta^2$ with $n_\ell = {\rm constant}$
provides a good approximation \cite{GGT-GW},
and then the solution of Eq. (\ref{GW-eq2}) is given by
\bea
   h_{\ell {\bf k}} (t)
       = {\sqrt{|\eta|} \over 
       a \sqrt{1 + \lambda_\ell \kappa^2 \dot \nu k/a}}
       \Big[ \tilde c_{1\ell} ({\bf k}) H_{\nu_\ell}^{(1)} (k|\eta|)
       + \tilde c_{2\ell} ({\bf k}) H_{\nu_\ell}^{(2)} (k|\eta|) \Big],
   \label{GW-exact-sol}
\eea
where $H_{\nu_\ell}^{(1)}$ and $H_{\nu_\ell}^{(2)}$ are Hankel functions
of the first and second kinds with
$\nu_\ell \equiv \sqrt{n_\ell +1/4}$.
Even for generic form of $z_\ell$,
we can still derive the {\it asymptotic} solution of Eq. (\ref{GW-eq2}).
In the small-scale limit where $k^2$ term dominates, we have a general 
solution given by
\bea
   h_{\ell {\bf k}} (t) = {1 \over 
       a \sqrt{ 1 + \kappa^2 \lambda_\ell \dot \nu k / a}} \Big[
       c_{1\ell} ({\bf k}) e^{i k \eta} 
       + c_{2\ell} ({\bf k}) e^{- i k \eta} \Big],
   \label{GW-SS-sol}
\eea
where $c_{1\ell}$ and $c_{2\ell}$ are integration constants of the left
and right travelling waves.
In the opposite limit of negligible $k^2$ term, 
which we call the large-scale limit,
we have a general solution of the form
\bea
   h_{\ell {\bf k}} (t) = C_\ell ({\bf k}) - D_\ell ({\bf k}) \int_0^t
       {dt \over a^3 ( 1 + \lambda_\ell \kappa^2 \dot \nu k/a )},
   \label{GW-LS-sol}
\eea
where $C_\ell$ and $D_\ell$ are the coefficients of relatively
growing and decaying modes, respectively.
Ignoring the transient solution, the above solution
manifestly showes that the amplitudes of both
polarization states of the tensor-type perturbation are conserved
in the large scale limit.
Notice that the asymptotic solutions in Eqs. (\ref{GW-SS-sol})
and (\ref{GW-LS-sol}) 
are valid for generic forms of 
time varying $V (\phi)$, $\omega (\phi)$, 
and $\nu (\phi)$.

Remarkably, the conservation properties in 
Eqs. (\ref{rot-sol}) and (\ref{GW-LS-sol}) are valid for 
generic forms of the axion potential $V(\phi)$, the axion
K\"ahler metric $\omega (\phi)$, and 
also  the axion coupling $\nu(\phi)$ to $R\tilde{R}$.
The large-scale conservation property of tensor-type perturbation
is particularly relevant in inflationary scenario.
During the transition period from inflation to ordinary radiation era,
the observationally relevant scales stay in the superhorizon size 
for which  the large-scale condition may apply.
As long as the scale remains in the large-scale limit, the conservation
property of tensor type perturbation
is valid  independently of how the axion field
(and also other scalar fields which may have 
non-minimal coupling to gravity) 
is settled down to its vacuum expectation value 
during the transition period.

Since it distinguishes the states
with different polarization as in Eq. (\ref{GW-eq}),
the axion coupling generically leads to  polarization asymmetry
in tensor-type perturbations.
In particular, if inflation  involves a period  
in which $\dot \nu k/a\sim 1/\kappa^2$,
the resulting polarization
asymmetry can be sizable.
Of course, this would not take place 
if the cosmic evolution during the whole inflation period
is determined  by the  dynamics
at energy scales
significantly below the Planck scale $M_P\equiv 1/\sqrt{\kappa}$.
However this can be a possibility
in string cosmology scenario  which encounters
a curvature  singularity during inflation at lowest order
approximation.
In such inflation models, higher dimensional
terms are expected to regulate  the curvature  singularity
to a value comparable to $1/\kappa^2$,
and then the axion coupling to $R\tilde{R}$
can give a  non-negligible effect although it is suppressed
by $\kappa^2$.
The evolution  of tensor-type perturbation
after inflation is  mainly described by
the large-scale solution (\ref{GW-LS-sol}) with conserved amplitudes.
Thus once the  polarization  asymmetry  
were generated during inflation, it will be preserved
over the subsequent evolution  as long as the scale remains
in the large-scale limit.
This then  may lead to an observable trace in the large angular scale 
polarization asymmetry  in the CMBR \cite{Lue}.
{}For a more  quantitive study of this problem, we will need specific 
inflation model for which our results can be easily applied.

\section*{Acknowledgments}
K. C and K. W. H.  are  supported in part by KOSEF
grant 98-0201-004-2, KRF Distinguished Scholar Exchange Program
and KRF  BSRI Program 1998-015-D00071.
J. H. wishes to thank Prof. Kamionkowski for informing us his work
during 19th Texas Symposium in Paris.



\begin{references}
\bibitem{string}
         M. Green, J. Schwarz and E. Witten, {\it Superstring Theory},
            Vols. 1 and 2 (Cambridge Univ. Press, Cambridge, 1987);
         J. Polchinski, {\it String Theory},
            Vols. I and II (Cambridge Univ. Press, Cambridge, 1998).
\bibitem{seehowever}	    
         I. Antoniadis, Phys. Lett. B {\bf 246}, 377 (1990);
	 I. Antoniadis, C. Munoz and M. Quiros, Nucl. Phys.
       	    B {\bf 397}, 515 (1993); 
	 J. D. Lykken, Phys. Rev. D {\bf 54}, 3693 (1996);
         N. Arkani-Hamed, S. Dimopoulos and G. Dvali, hep-ph/9803315;
	 I. Antoniadis, N. Arkani-Hamed, S. Dimopoulos and G. Dvali,
      	    hep-ph/9804398; 
	 C. P. Burgess, L. E. Ibanez and F. Quevedo, hep-ph/9810535.
\bibitem{pre-big-bang}
         G. Veneziano, Phys. Lett. B {\bf 265}, 287 (1991);
	 M. Gasperini and G. Veneziano, Aspropart. Phys.
	 {\bf 1}, 317 (1993);
	 R. Brustein,  M. Gasperini, M. Giovannini and G. Veneziano,
	 Phys. Lett. B {\bf 361}, 45 (1995);
         J. Hwang, Astropart. Phys. {\bf 8}, 201 (1998);
         R. Brustein, hep-th/9801008.	 
\bibitem{string-action}
         C. G. Callan, D. Friedan, E. J. Martinec and M. J. Perry, 
	       Nucl. Phys. B {\bf 262} 593 (1985);
         B. Zwiebach, Phys. Lett. B {\bf 156}, 315 (1985);
         S. Deser and A. N. Redlich, {\it ibid.} {\bf 176}, 350 (1986);
         D. J. Gross and J. H. Sloan, Nucl. Phys. B {\bf 291}, 41 (1987).
\bibitem{Antoniadis}
         I. Antoniadis, J. Rizos, and K. Tamvakis, Nucl. Phys. B
            {\bf 415}, 497 (1994);
	 R. Brustein and R. Madden hep-th/9901044.   
\bibitem{stringaxion}
         E. Witten, Phys. Lett. B {\bf 149}, 351 (1984); 
	 K. Choi and J. E. Kim, {\it ibid.} {\bf 154}, 393 (1985);
 	 {\it ibid.} {\bf 165}, 71 (1985).
\bibitem{Maxion}
         T. Banks and M. Dine, Nucl. Phys. B {\bf 479}, 173 (1996);
 	 K. Choi, Phys. Rev. D {\bf 56}, 6588 (1997).
\bibitem{ram} For cosmological implications of  
       the axion coupling to $F\tilde{F}$, see
        R. Brustein and D. H. Oaknin, Phys. Rev. Lett. {\bf 82},
	2628 (1999); Phys. Rev. D {\bf 60}, 023508 (1999).

\bibitem{Lue}
         A. Lue, L. Wang, and M. Kamionkowski,  astro-ph/9812088.
\bibitem{cosmo-appl}
         R. Easther and K. Maeda, Phys. Rev. D {\bf 54}, 7252 (1996);
         P. Kanti, J. Rizos and K. Tamvakis, gr-qc/9806085.
\bibitem{Gasperini}
         M. Gasperini, Phys. Rev. D {\bf 56}, 4815 (1997).
\bibitem{Kawai}
         S. Kawai, M. Sakagami and J. Soda, Phys. Lett. B {\bf 437}, 284 (1998).
\bibitem{GB}
         J. Hwang, Phys. Rev. D {\bf 53}, 762 (1996);
         J. Hwang and H. Noh, {\it ibid.} in press (1999).
\bibitem{Rab-GW}
         H. Noh and J. Hwang, Phys. Rev. D {\bf 55}, 5222 (1997).
\bibitem{Rab-rot}
         J. Hwang and H. Noh, Phys. Rev. D {\bf 57}, 2617 (1998).
\bibitem{GGT-GW}
         J. Hwang, Class. Quant. Grav. {\bf 15}, 1401 (1998).
\end{references}
\end{document}